\documentclass[12pt,a4paper]{article}
\usepackage{authblk}
\usepackage[T1]{fontenc}
\usepackage[latin1]{inputenc}
\usepackage{textcomp}
\usepackage{amssymb}
\usepackage{amsmath}
\usepackage{bm,upgreek}
\usepackage{graphicx}
\graphicspath{{./figures/}}
\usepackage{hyperref}
\usepackage{rotating}
\usepackage{pdflscape}
\usepackage{lineno}
\usepackage{multirow}
\usepackage{array}
\usepackage{algorithm}
\usepackage{algpseudocode}
\usepackage{xcolor}
\usepackage{soul}
\usepackage{natbib}
\usepackage{calrsfs}
\usepackage{animate}
\usepackage{color}
\usepackage{tabularx}
\usepackage{textgreek}
\usepackage{setspace}
\usepackage[top=22mm, bottom=22mm, left=22mm, right=22mm]{geometry}

\usepackage{float}
\floatstyle{plaintop}
\restylefloat{table}
\usepackage[tableposition=top]{caption}
\newcommand{\crm}[1]{\romannumeral #1}
\renewcommand{\contentsname}{Contents \color{red}- will be removed before submission\color{black}}

\begin{document}

\title{Reduction of conceptual model uncertainty using ground-penetrating radar profiles: Field-demonstration for a braided-river aquifer}

\author[1]{Guillaume Pirot}
\author[2,3]{Emanuel Huber}
\author[1]{James Irving}
\author[1]{Niklas Linde}
\affil[1]{Applied and Environmental Geophysics Group, Institute of Earth Sciences (ISTE), University of Lausanne, Switzerland}
\affil[2]{Department of Geological Sciences, Stanford University, USA}
\affil[3]{Applied and Environmental Geology, University of Basel, Switzerland}

\date{December 13, 2018}
\maketitle

% \section*{Keywords}

% \section*{Abstract}

% \tableofcontents
% \clearpage
% \linenumbers
\section*{Abstract}
Hydrogeological flow and transport strongly depend on the connectivity of subsurface properties. Uncertainty concerning the underlying geological setting, due to a lack of field data and prior knowledge, calls for an evaluation of alternative geological conceptual models. To reduce the computational costs associated with inversions (parameter estimation for a given conceptual model), it is beneficial to rank and discard unlikely conceptual models prior to inversion. Here, we demonstrate an approach based on a quantitative comparison of ground-penetrating radar (GPR) sections obtained from field data with corresponding simulation results arising from various geological scenarios. The comparison is based on three global distance measures related to wavelet decomposition, multiple-point histograms, and connectivity that capture geometrical characteristics of geophysical reflection images. Using field data from the Tagliamento braided river system, Italy, we demonstrate that seven out of nine considered geological scenarios can be discarded as they produce GPR sections that are incompatible with those observed in the field. 
The retained scenarios reproduce important features such as cross-stratified deposits and irregular property interfaces. The most convenient distance measure of those considered is the one based on wavelet-decomposition. Direct analysis of the distances  is the most intuitive and fastest way to compare scenarios.

\section{Introduction}
Reliable predictions of groundwater flow and contaminant transport require adequate characterization of subsurface properties and their connectivity \citep[e.g.,][]{gomezHernandez1998,zinn2003}.
In this regard, limited number of data and knowledge of the field site implies that multiple geological conceptual models must be initially considered. That is to say models with different geometrical characteristics of the deposits, such as channels, lenses or layers.
%Since connectivity of high and low-permeability units has a strong impact on predictions, and because field knowledge is limited, several possible geological conceptual models should be considered that describe the spatial organization of property fields (e.g. geometrical characteristics of the deposits, such as channels, lenses or layers). 
A general approach to compare alternative geological conceptual models is to perform Bayesian model selection based on field data acquisition and inversion. 
It aims at estimating the Bayes factors, that is, the ratios of the estimated evidences (i.e., the integral of the likelihood over the prior probability density function) for the considered scenarios \citep{kass1995,schoniger2014}. %[see][for a review]
However, reliable evidence estimators are costly because they necessitate a very large number of numerical evaluations of property models. As a result, modelers often assume a single conceptual model \citep{ferre2017} on which they perform inversion on the distribution of physical properties such as hydraulic conductivity, porosity or storativity \citep{carrera1986a,hojberg2005,eaton2006} for a given geological conceptual model. 
The main risks associated with such a practice is underestimation of uncertainty and  biased parameter distributions and predictions. There is, thus, a need for efficient, albeit more approximate, ways to compare alternative geological scenarios without resorting to formal evidence computations. \newline

To enable comparison of geological conceptual models using a reduced number of costly forward simulations, \cite{park2013} draw property models from each of the considered scenarios and calculate their data response. 
They then use multi-dimensional scaling (MDS) to reduce dimensionality, followed by adaptive kernel smoothing to estimate the probability of each scenario by comparing its distance to the reference data. Sometimes, it can be beneficial to base such comparisons on data types other than classical hydrogeological data \citep{huber2016}.
Non-invasive geophysical data, for example, can provide substantial information about connectivity, structure dimensions and orientations, and thus might help to reduce geological conceptual model uncertainty. 
%Indeed, seismic, ground-penetrating radar (GPR), electric resistivity tomography (ERT) or electromagnetic (EM) surveys, for instance, are generally non-invasive and sensitive to subsurface property variations. 
Notably, geophysical images reflect the sensitivity of the employed method to subsurface property variations. Thus, they can provide information about length scales and orientation characteristics of significant property boundaries. The wide range of available geophysical techniques offer flexibility to adjust resolution or depth of investigation, and  to maximize the sensitivity to subsurface properties of interest \citep{hubbard2005}.
For instance, comparisons of seismic images \citep{scheidt2015} or of electric resistivity tomography (ERT) images \citep{hermans2015} offer possibilities to falsify scenarios or reduce conceptual model uncertainty.\newline

Possibly the simplest way to quantitatively compare geophysical images is to use a distance based on pixelwise (one-to-one) local comparison \citep{hermans2015}. %, whereby the two compared images of identical size are subtracted and applied a $p$ norm. %[as performed by][in their comparison of ERT profiles]
 However, by using a local comparison, the probability of sharing a majority of similar pixel values and thus to observe small distances is quite low. So, when the main interest lies in the comparison of patterns and not the specific locations of property values, approaches relying on global geometrical characteristics are better suited. Approaches to sort and classify images in this way has been widely studied in the field of image processing \citep{smeulders2000}. Among many alternatives, those based on discrete wavelet transforms have proven efficient to identify the images that are the closest in a large database. \cite{suzuki2008} and \cite{scheidt2009} use a distance based on wavelet decomposition \citep{mallat1989} of geological realizations for different scenarios to represent spatial uncertainty. \cite{scheidt2015} further apply this type of metric on seismic images to update probabilities of alternative prior scenarios. Nevertheless, distances based on wavelet decomposition rely on the comparison of coefficient histograms, which might hide spatial characteristics such as pattern connectivity. It is, thus, important to also consider other distances, for instance, based on multiple-point histogram \citep{boisvert2010} or connectivity analysis \citep{renard2013,meerschman2013}, that allow quantitative comparison of the global spatial characteristics of interest obtained from field data with those obtained from synthetic modeling based on various scenarios.\newline
% * <ehuber@stanford.edu> 2018-09-07T08:10:00.454Z:
% 
% >  the probability of sharing a majority of similar pixel values and thus to observe small distances is quite low.
% Proposition:
% the probability to observe similar pixel values between two data and thus to obtain small distances is quite low.
% 
% ^.

So far and to the best of our knowledge, quantitative approaches to reduce conceptual geological model uncertainty using image comparisons did not consider multiple distance types and there has been no such application to GPR data. Traditionally, GPR data are interpreted qualitatively and its quantitative integration in subsurface modeling is largely unexplored. %And, as mentioned before, \cite{scheidt2015} compared seismic images based on a wavelet-decomposition distance and \cite{hermans2015} compared ERT images using a pixelwise distance. 
In the continuity of previous related works \citep{park2013,pirot2014b,scheidt2015,hermans2015}, we propose to extend such approaches to GPR reflection sections, using different distance measures of global geometrical characteristics. The three types of distances considered herein for the comparison of GPR reflection sections are based on 1) wavelet decomposition, 2) multiple-point histogram and 3) connectivity functions. 
In addition, the computed distances are analyzed and interpreted with a simple intuitive approach and with a more complex formal approach based on dimensionality reduction and mapping techniques. 
\newline

The objectives of this work are \crm{1}) to demonstrate how a simple but robust method enables the comparison of global characteristics of GPR reflection sections obtained from field data processing with those obtained from GPR reflection sections simulated from different scenario realizations; \crm{2}) to verify that GPR reflection sections can be used to reduce geological conceptual model uncertainty; \crm{3}) to investigate the relative strengths of three different distance measures for GPR data; and \crm{4}) to present follow-up strategies depending on the closeness or remoteness of simulated sections with reference sections obtained from field measurements. 
To illustrate the proposed method, we consider GPR profiles acquired on the riverbed of the Tagliamento River, Northeast Italy \citep{huber2015phd}. We consider three different geological conceptual models; each one of them being sub-divided in three sets of parameters (scenarios). For each of the nine resulting scenarios, 20 stochastic aquifer realizations are used as inputs for GPR simulations.
The distances are used to produce a first ranking and to falsify unlikely scenarios. A dimension reduction technique called multi-dimensional scaling (MDS) followed by kernel smoothing are then used to estimate scenario probabilities. 
\newline

The paper is organized as follows. Section \ref{secDistances} describes the distance measures considered and how they can be used to update scenario probabilities. Section \ref{secApplication} presents a field-demonstration using GPR sections simulated from realizations of different geological conceptual geological models of the Tagliamento site (subsection \ref{subsecGeologicalPrior}). This section continues with the presentation of the migrated field GPR data and its processing steps (subsection \ref{subsecGPRprocessing}), and ends with the simulation of migrated GPR profiles (subsection \ref{subsecGPRsim}). Section \ref{secResults} displays the results, which are further discussed in Section \ref{secDiscussion}. Conclusions are given in Section \ref{secConclusion}. %\newline

\section{Distances between geophysical images and estimation of scenario probabilities}
\label{secDistances}
In this section, we briefly review three distance measures that can be used to compare global geometrical characteristics of geophysical images. We then describe how approximate scenario probabilities can be obtained from field and simulated data through MDS and adaptive kernel smoothing \citep{park2013}. 

\subsection{Wavelet decomposition}
One way to extract global characteristics of an image is wavelet decomposition \citep{mallat1989}. We consider in our work the same decomposition as \cite{scheidt2015}. Two geophysical images $i_1$ and $i_2$ are decomposed in two levels by a ``Haar'' wavelet
%[a square shape function defined as 1 inside $\left[0;1/2\right[$, as -1 inside $\left[1/2;1\right[$ and equal to 0 elsewhere,][]
\citep{haar1910}, which produces a series of coefficients (horizontal, vertical, diagonal and approximation) for each level. At each level, the histogram of each coefficient is discretized into bins $b\in 1 \dots B$, using the same binning for both images. For each level $m\in 1 \dots M$ and each coefficient $c\in 1 \dots C$, a distance $d_{JS}$ between the two images is computed based on the Jensen-Shannon divergence between the probability distributions $P_1^{m,c}$ and $P_2^{m,c}$ derived from these histograms:
\begin{equation}
d_{JS}(i_1,i_2,m,c) =  \frac{ d_{KL}( P_1^{m,c},\frac{P_1^{m,c}+P_2^{m,c}}{2} ) + d_{KL}( P_2^{m,c},\frac{P_1^{m,c}+P_2^{m,c}}{2} ) }{2}  \textnormal{,}
\label{GPeqdjs}
\end{equation}
where $d_{KL}(P,Q)$ is the Kullback-Leibler divergence between discrete probability distributions $P$ and $Q$ computed as $d_{KL}(P,Q)=\sum\limits_{b=1}^B P(b) \log \frac{Q(b)}{P(b)}$ \citep{kullback1951}.
Then, the corresponding wavelet-based distance $D_w(i_1,i_2)$ is:
\begin{equation}
D_w(i_1,i_2) =   \sum_{m=1}^M \sum_{c=1}^C \frac{d_{JS}(i_1,i_2,m,c)}{M \times C} \textnormal{.}
\label{GPeqDw}
\end{equation}

\subsection{Multiple-point histogram}
Another way to quantify global spatial characteristics of an image is to define a summary statistic describing its multiple-point histogram \citep{boisvert2010}. %[see][for more details]
In multiple-point statistics (MPS), a pattern is usually defined as a set of values associated with relative coordinates that define a spatial configuration. Two patterns are distinct when the values are different at one of the relative coordinates. The multiple-point histogram (MPH) of an image is defined for a given spatial configuration, also called search window, as the occurrence list of distinct patterns. Here we use the \emph{Impala} \citep{straubhaar2013} software to compute multiple-point histograms from categorical geophysical images. Note, however, that the measure can be adapted to deal with continuous geophysical images (see Section \ref{subsecDiscussionMeas}).
%We use a pattern geometry given by a $5 \times 5$ pixel window. 
Multiple-point histograms are computed at $M$ multigrid levels $m$, to account for patterns at, relatively speaking, small, intermediate and large scales \citep{tran1994}. A multigrid is practical to account for larger scale structures while keeping the pattern geometry and, thus, the computing time reasonable. Each histogram is limited to the $O$ most frequent patterns $o$. By denoting $f_{i}^{o,m}$ the frequency of pattern $o$ at level $m$ in image $i$, the multiple-point histogram based distance $D_{mph}$ between image $i_1$ and image $i_2$ is defined as:
\begin{equation}
D_{mph}(i_1,i_2) =   \sum_{m=1}^M \sum_{o=1}^O \frac{|f_{i_1}^{o,m} - f_{i_2}^{o,m}| \times (f_{i_1}^{o,m} + f_{i_2}^{o,m})}{2 \times M \times O} \textnormal{.}
\label{GPeqDmph}
\end{equation}

\subsection{Connectivity measure}
The final measure that we consider to quantify global characteristics of an image is connectivity \citep{renard2013}. Indeed, subsurface property connectivity dictates subsurface flow paths and transport. Here we consider categorical geophysical images, but note, that the measure can be adapted to deal with continuous geophysical images \citep{pirot2014}. We consider connectivity as the probability that two pixels belonging to the same class (a range of values) are connected, as a function of the distance and direction, similarly to the definition of a directional semi-variogram \citep{matheron1963}.
By denoting $C(i,a,l)$ the connectivity measure of a discrete image $i$ along axis $a \in 1 \dots A$ for a distance lag $l \in 1 \dots L$, the connectivity distance $D_c(i_1,i_2)$ between discrete images $i_1$ and $i_2$ can be computed \citep{meerschman2013} as 
\begin{equation}
D_c(i_1,i_2) = \sum_{a=1}^A \sum_{l=1}^L \frac{ | C(i_1,a,l) - C(i_2,a,l) |}{A \times L}\  \textnormal{.}
\label{GPeqDc}
\end{equation}

\subsection{Estimation of scenario probabilities}
To assess the probability of a scenario given a geophysical section, we follow the approach by \cite{park2013}.
Given a distance metric $D$ and an ensemble of $I$ images $i$, the distance between all pairs $i_j,i_k$ of images define a dissimilarity matrix  $\delta_{jk}=D(i_j,i_k)$. Multidimensional scaling \citep[MDS,][]{cox2000} is a method to represent the images as points in a low dimensional space, usually Euclidean. While principal component analysis (PCA) requires point coordinates, MDS can be used on data for which only the relative distances are known. This lower dimensional space is searched, such that the distances $d_{jk}$ between the points are as close as possible to the original dissimilarity matrix  $\delta_{jk}$. 
MDS allows to map images in space, as points, for instance in 2D if using the two main dimensions. Now, we consider reference points related to reference images and a cloud of points related to images derived from a scenario. We can approximate the density of the cloud at any location of the low dimensional space, using adaptive kernel smoothing \citep{ebeling2006}. For each scenario $s$, the density at one or several reference points (in the low dimensional space) can be computed as a scalar $\rho_s$. The updated probability $P$ of scenario $s$ can then be approximated as $P(s)=\frac{\rho(s)}{\sum_s \rho(s)}$. These updated probabilities are relative to the ensemble of considered scenarios, with $P(s)$ the probability that an image generated from scenario $s$ is the closest to the reference image.

\section{Field application and GPR modeling}
\label{secApplication}

\begin{figure}[ht!]
	\centering
	\includegraphics[width=.5\textwidth]{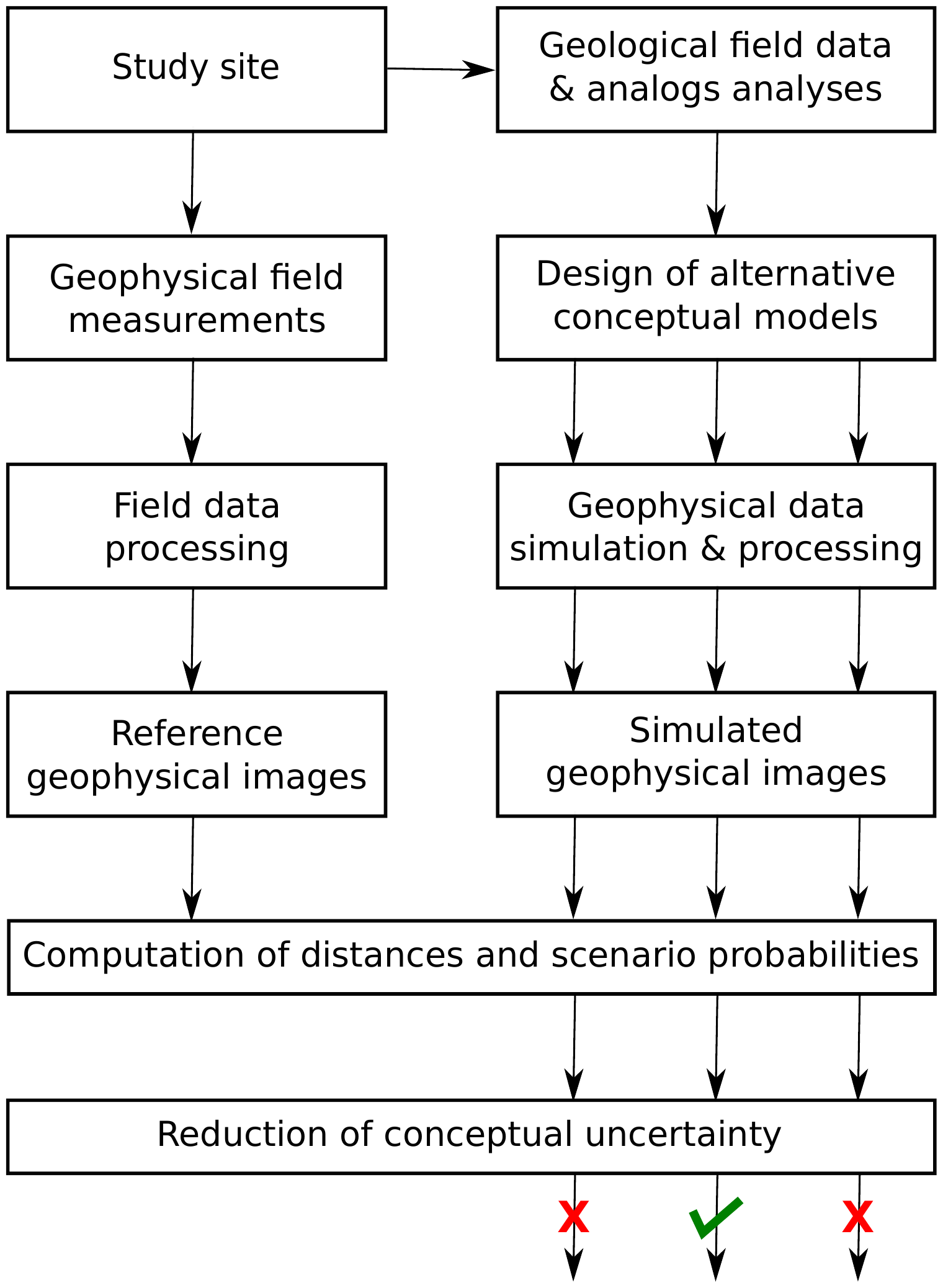}
	\caption{Overview of the workflow to reduce geological conceptual model uncertainty. On the left, the path of arrows represents field data processing; on the right, the three vertical arrow paths represents the workflow for three distinct scenarios; at the bottom, a red cross illustrates scenario falsification and a green mark indicates scenario compatibility.}
	\label{GPfig0}
\end{figure}

A pre-requisite to compare field and simulated data (Figure \ref{GPfig0}) is to apply equivalent data processing \citep{hermans2015}, but this is rarely sufficient because actual field conditions always differ from numerical implementations. Indeed, results obtained from the processing of geophysical data are prone to errors \citep[e.g.,][]{linde2014} related to field data acquisition, simplifications in physical modeling or consequences of numerical modeling such as numerical and geometrical approximations. For instance, seismic or GPR geophysical images obtained from field data might include false discontinuities and their interpretation in terms of continuous connected structures or interface delineation necessitates expert knowledge. On the contrary, seismic or GPR geophysical images obtained from forward modeling, might reproduce property (dis)continuities too well and appear too clean to be representative of what would be expected for real data. To further reduce the remaining gaps between the results obtained from field data and from synthetic scenarios, it is necessary to include fit for purpose filtering \citep{green1988,panagiotakis2011} such that geophysical sections are not dominated by details/aspects that we do not seek to reproduce. 

\subsection{Study site and geological conceptual models}
\label{subsecGeologicalPrior}
\begin{figure}[ht!]
	\centering
	\includegraphics[width=.9\textwidth]{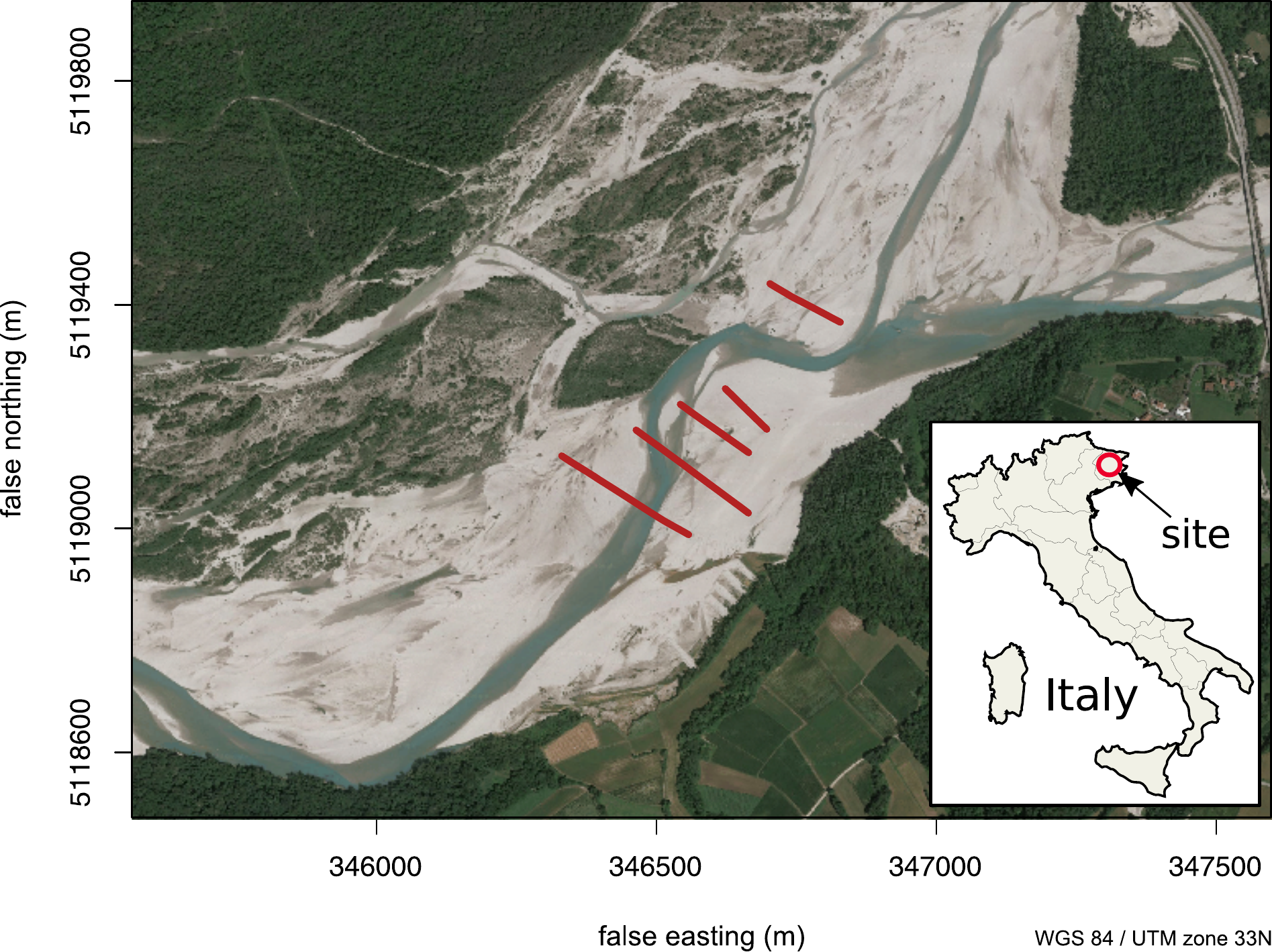}
	\caption{Site location in Italy (map from \url{http://www.pedagogie.ac-aix-marseille.fr/jcms/c_67064/en/cartotheque}); position of the GPR profiles over an aerial photograph of the Tagliamento river, south east of Flagogna (Google maps satellite image).}
	\label{GPfig0b}
\end{figure}

The study site considered is a portion of a sandy-gravel aquifer located near the city of Flagogna, Italy, within a portion of the active bed of the gravelly braided Tagliamento river (Figure \ref{GPfig0b}). 
%\begin{figure}[ht!]
%	\centering
%	\includegraphics[width=.9\textwidth]{localization}
%	\caption{(a) map of Italy (from \url{http://www.pedagogie.ac-aix-marseille.fr/jcms/c_67064/en/cartotheque}), the red star indicates the location of the study site close to the city of Flagogna; (b) aerial photograph of the Tagliamento river, south east of Flagogna (Google maps satellite image).}
%	\label{GPfig0b}
%\end{figure}
The Tagliamento river flows in the Friuli Venezia Giulia region, northeastern Italy, from the Carnian Alps to the Adriatic Sea. As the Tagliamento river is one of the few remaining large semi-natural rivers in the Alps \citep{ward1999} it was chosen as a study site to characterize the link between the topography of the active river bed and subsurface properties \citep{huber2015}. GPR data acquisitions  and interpretations allowed to improve the characterization of scours and to model them \citep{huber2016b}. In addition to improving the understanding of deposition and erosion processes \citep{huber2016}, this work inspired modelers to develop new methods, such as a pseudo-genetic approach to produce heterogeneous models of braided-river aquifers \citep{pirot2015a}. \newline

\begin{figure}[ht!]
	\centering
	\includegraphics[width=1\textwidth]{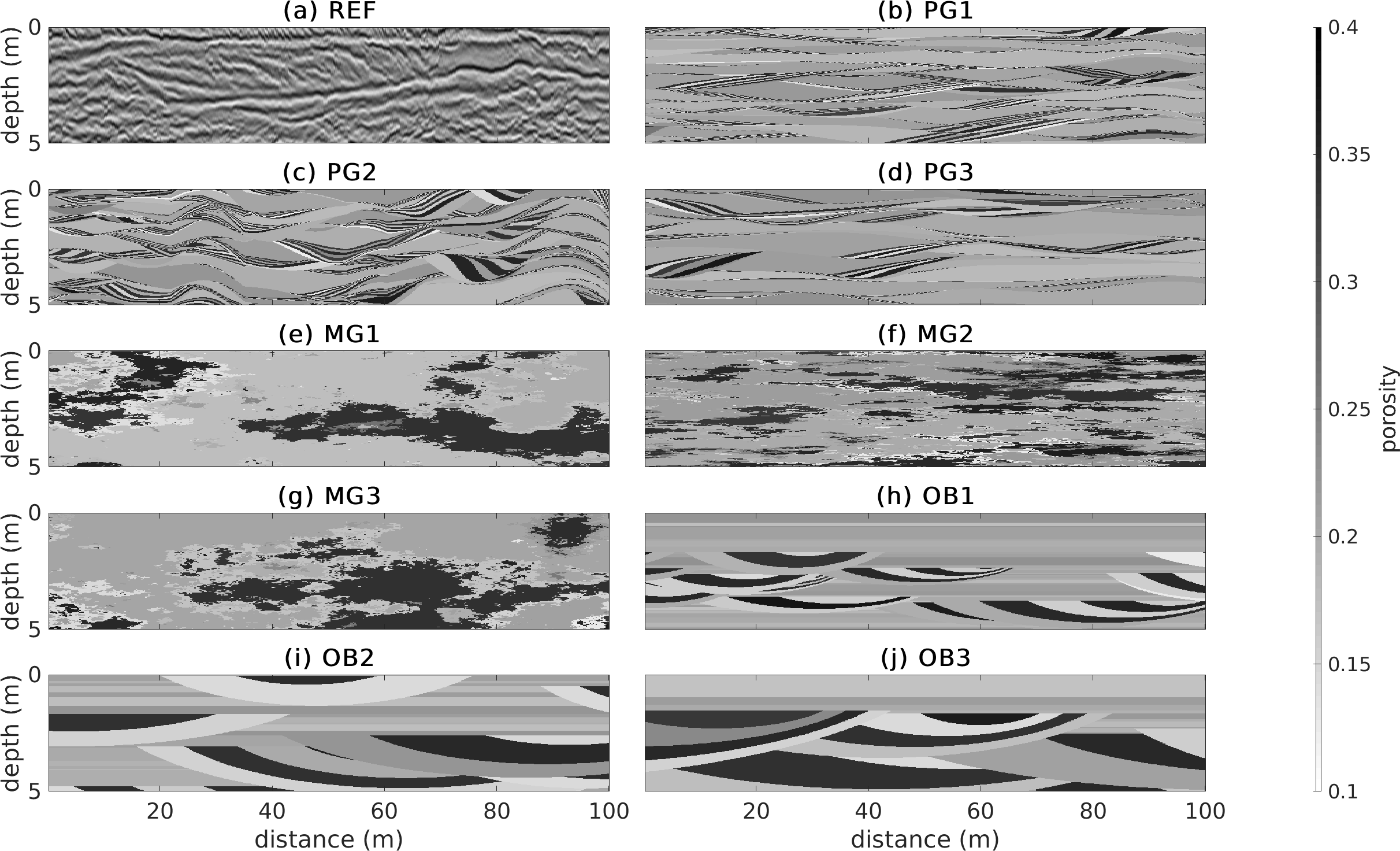}
	\caption{Example porosity sections for different geological scenarios that are to be compared to (a) a reference GPR reflection section processed from field data (REF01); (b), (c) \& (d) porosity sections from pseudo-genetic model realizations for parameter sets PG1, PG2 \& PG3, respectively; (e), (f) \& (g) porosity sections from truncated multi-Gaussian model realizations for parameter sets MG1, MG2 \& MG3, respectively; (h), (i) \& (j) porosity sections from object-based model realizations for parameter sets OB1, OB2 \& OB3, respectively.}
	\label{GPfig1}
\end{figure}

Assuming a braided-river type of aquifer, we wish to investigate which geological conceptual model is best suited to represent the porosity field. To this end, we consider subsets of reflection GPR sections in the saturated zone. Indeed, below the water table, GPR responses are strongly dependent on the porosity variations in the subsurface \citep{daniels2004}. 
We consider three different types of conceptual models of porosity , similar to those considered by \citet{pirot2015b} in their assessment of the impact of geological conceptual models on contaminant transport. Each type of geological conceptual porosity model is sub-divided into three sets of parameter values (scenarios) with geometrical features (patterns) that present different length scales (Figure \ref{GPfig1}). 
Here we further assume that the braided-river aquifer is composed of three structural elements: gray gravel (GG), bimodal (BM) and open-framework (OW) deposits. Each distinguishable geobody or sedimentary deposit is a assigned a randomly drawn value from the  porosity distribution, related to its structural element \citep{jussel1994}, as described in Table \ref{GPtab1}. The models are characterized by a horizontal discretization of $0.25\:m$ and a vertical discretization of $0.01\:m$.\newline

\begin{table}[ht!]
	%\scriptsize
	\small
    \centering
	\begin{tabular}{l c c c}
		\hline
		Structural Element & Pdf Law & Porosity Mean ($\%$) & Porosity Standard Deviation ($\%$) \\
		\hline
		GG & normal & 20.1 & 1.4 \\
		BM & normal & 18.8 & 3.9 \\
		OW & normal & 34.9 & 1.4 \\
		\hline
	\end{tabular}
	\normalsize
	\caption{Probability density function (pdf) properties of the porosity for each structural element \citep[from][]{jussel1994}.}
	\label{GPtab1}
\end{table}

The first geological geological conceptual model is represented by realizations from a pseudo-genetic (PG) algorithm \citep{pirot2015a}, which mimics deposition and erosion steps by stacking successive simulated topographies, and by imitating sandy-gravel material transport and deposition. Here, the main layers are populated with GG elements and the resulting cross-stratified deposits by successive BM and OW elements. A second geological geological conceptual model is a truncated multi-Gaussian (MG) model \citep{emery2006}, in which the locations above the highest threshold are populated with OW elements, the location between the two thresholds are defined as BM elements, and the remaining matrix is populated with GG elements. 
The third geological geological conceptual model is an object-based (OB) model \citep{huber2016b} mathematically defined as a compound marked Strauss process. The OB simulates the formation of spoon-shaped structures on the river bed and the subsequent deposition of sediments over the whole river bed. The spoon-shaped structures are modeled by truncated ellipsoids with an internal OW--BM cross-bedding and the sediments deposited on the river bed by horizontal layers of GG \citep[e.g.,][]{beres1999, huggenberger2006}. The parameters underlying each scenario are summarized in Table \ref{GPtab2}; they were chosen to approximate the dimensions of scours that were estimated from field observations and from interpretations of migrated GPR sections.

\begin{table}[ht!]
	%\scriptsize
	\small
    \centering
	\begin{tabular}{l c c c c c c c}
		\hline
		Scenario & Example & \multicolumn{5}{c}{Parameters} \\
		\hline
		& & Scalability & Scalability & Aggradation & Number of & Deposition \\
		& & Width & Depth & Range (m) & Iterations & Intensity \\
		\hline
		PG1 & Figure  \ref{GPfig1}b & 1 & 1 & $[0.05\:;\:0.125]$ & 8 & 5 \\
		PG2 & Figure  \ref{GPfig1}c & $1/2$ & $1.6$ & $[0.05\:;\:0.125]$ & 8 & 5 \\
		PG3 & Figure  \ref{GPfig1}d & $1/3$ & 1 & $[0.2\:;\:0.25]$ & 6 & 3.5 \vspace{2mm}\\
		\hline
		& & Variogram & Horizontal & Vertical & OW element & BM element \\
		& & Model & Range (m) & Range (m) & Proportion & Proportion \\
		\hline
		MG1 & Figure  \ref{GPfig1}e & exponential & 50 & 3 & 25\% & 25\% \\
		MG2 & Figure  \ref{GPfig1}f & exponential & 25 & 0.5 & 25\% & 25\% \\
		MG3 & Figure  \ref{GPfig1}g & exponential & 70 & 5 & 25\% & 25\% \vspace{2mm}\\
		\hline
		& & Width & Width/height & Layer Poisson & \multicolumn{2}{c}{Horizontal Strauss process} \\
		& & Range (m) & Ratio & Process ($\lambda$) & $\beta$ & $\gamma$ \\
		\hline
		OB1 & Figure  \ref{GPfig1}h & $[10\:;\:20]$ & $[11\:;\:18]$ & 0.1 & $10^{-3}$ & 0.5 & \\
		OB2 & Figure  \ref{GPfig1}i & $[22\:;\:33]$ & $[11\:;\:18]$ & 0.1 & $5 \: 10^{-4}$ & 1 & \\
		OB3 & Figure  \ref{GPfig1}j & $[35\:;\:53]$ & $[11\:;\:18]$ & 0.1 & $2.5\:10^{-4}$ & 1 & \\
		\hline
	\end{tabular}
	\normalsize
	\caption{Parameter choices for each scenario grouped by type of geological conceptual model.}
	\label{GPtab2}
\end{table}

\subsection{GPR data acquisition and processing}
\label{subsecGPRprocessing}
The reflections in the processed and migrated GPR sections provide indirect information about characteristic geometric features. Such sections are used herein to compare, based on various global distance measures, different types of geological conceptual models.
% Detailed steps:
% *** Class GPR ***
%  survey date =  2012-10-02 
%  Reflection, 100MHz,Window length=8m, dz=0.01m
%  1055 traces,263.5m long
%  PROCESSING
%    1. coord<-
%    2. crs<-
%    3. dcshift:1-50+mean
%    4. time0<-
%    5. timeCorOffset
%    6. dewow:type=MAD+w=30
%    7. fFilter:f=5,10,170,250+type=bandpass+plotSpec=NULL
%    8. gain:type=power+alpha=1+tcst=10+t0=0+te=200
%    9. gain:type=exp+alpha=0.015+t0=20+te=140
%    10. dewow:type=Gaussian+w=10
%    11. upsample:n=3+n=1
%    12. migration:type=kirchhoff+max_depth=8+dz=0.01+fdo=80+FUN=NULL
%    13. filter1D:type=Gaussian+sigma=2.5
%    14. gain:type=agc+w=0.45
%  ****************
Five GPR profiles (REF01 to REF05) were acquired on the Tagliamento riverbed, orthogonally to the main flow direction. REF01 section is used for comparison with simulated data, while REF02 to REF05 are used to assess on-site data variability. 
The GPR data were acquired with a PulseEkko Pro GPR system (Sensors \& Software Inc., Mississauga, Canada) using 100 MHz antennas and a measurement spacing of $0.25$ m. A common mid-point (CMP) was performed to estimate the mean GPR velocity.
The data processing steps are described in Table \ref{GPtab3} and they were carried out with the RGPR package \citep{RGPR}.
\begin{table}[ht!]
	\small
	\centering
	\begin{tabular}[t]{r p{15cm}}
		\hline
		Step & Description \\
		\hline
	 	1 & DC-shift \\
	 	2 & time zero correction \\ 
	 	3 & dewow to remove the low frequency trend in the signal \\
	 	4 & band pass filter to remove noise ($7< \textnormal{signal} <200$ MHz, defined as a stepwise linear function between, 5,10,170 \& 250 MHz) \\
	 	5 & power gain \& exponential gain ($\alpha = 1$) to correct for geometric spreading and attenuation depth \citep{kruse2003,grimm2006} \\
	 	6 & dewow to correct for the deviation from zero that is reinforced by the power and exponential gains \\
	 	7 & topographic Kirchhoff migration with a constant velocity $\overline{vel}=100 \: $m/\textmu s\\
	 	8 & 1D vertical Gaussian (standard deviation $\sigma = 2.5$ cm) low-pass filter to lightly smooth the migrated image and get rid of persisting high frequency noise \\
	 	9 & automatic gain control to balance signal amplitudes \citep[standard deviation of the Gaussian filter $\sigma=0.45\:$m, power used to compute the p-norm $p=2$ \& $r=1/p$; see][for more details]{rajagopalan1994}\\
		\hline
	\end{tabular}
	\normalsize
	\caption{Processing steps applied to field GPR reflection data.}
	\label{GPtab3}
\end{table}
The migrated section corresponding to the REF01 profile is presented in Figure \ref{GPfig1}a.
\begin{figure}[ht!]
	\centering
	\includegraphics[width=1\textwidth]{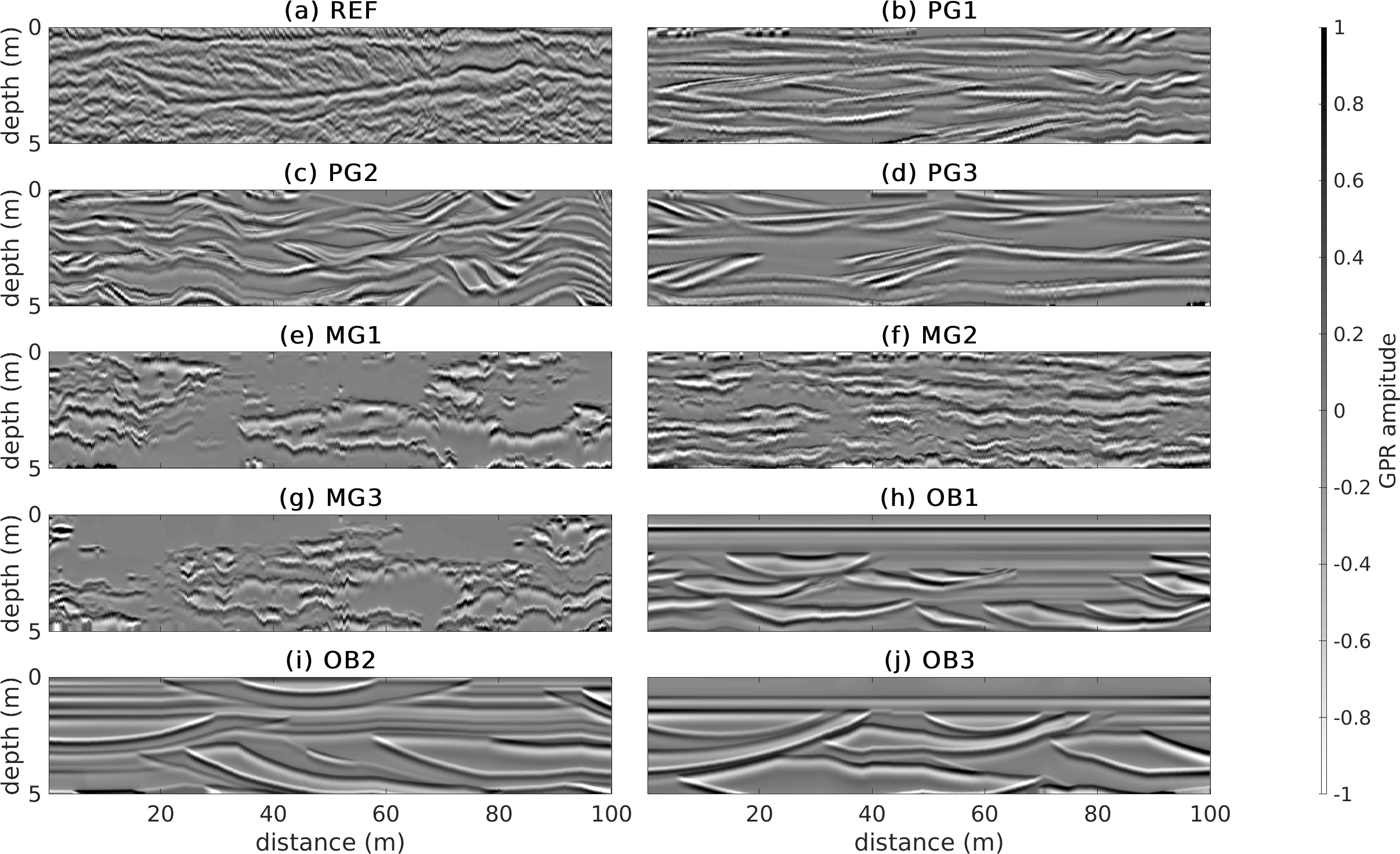}
	\caption{(a) Processed and migrated GPR reflection section from field data (REF01); (b), (c) \& (d) GPR reflection sections simulated from pseudo-genetic porosity model realizations for parameter sets PG1, PG2 \& PG3, respectively; (e), (f) \& (g) GPR reflection sections simulated from truncated multi-Gaussian porosity model realizations for parameter sets MG1, MG2 \& MG3, respectively; (h), (i) \& (j) GPR reflection sections simulated from object-based porosity model realizations for parameter sets OB1, OB2 \& OB3, respectively.}
	\label{GPfig2}
\end{figure}

The processed migrated sections are thresholded into binary images to focus on the predominant aspects of the reflections. The amplitude of the processed GPR reflection section is similar throughout the image after applying the automatic gain control. Consequently, at all interfaces where porosity changes, the signal amplitude is similar, independently of the porosity contrast. 
We consider the first (negative) and last (positive) quartiles of the signal amplitude in the section. We retain the last quartile of the reflections (positive amplitude) to define Class 1. Tests (not shown) indicated that it was not necessary to retain the first quartile (negative amplitude) to define another class, as the corresponding class would have almost the same geometrical characteristics as those of Class 1. Therefore, we use amplitudes below the $75^{th}$ percentile to define Class 2 (Figure \ref{GPfig3}a).

\subsection{From aquifer porosity models to GPR reflection sections}
\label{subsecGPRsim}
\begin{figure}[ht!]
	\centering
	\includegraphics[width=1\textwidth]{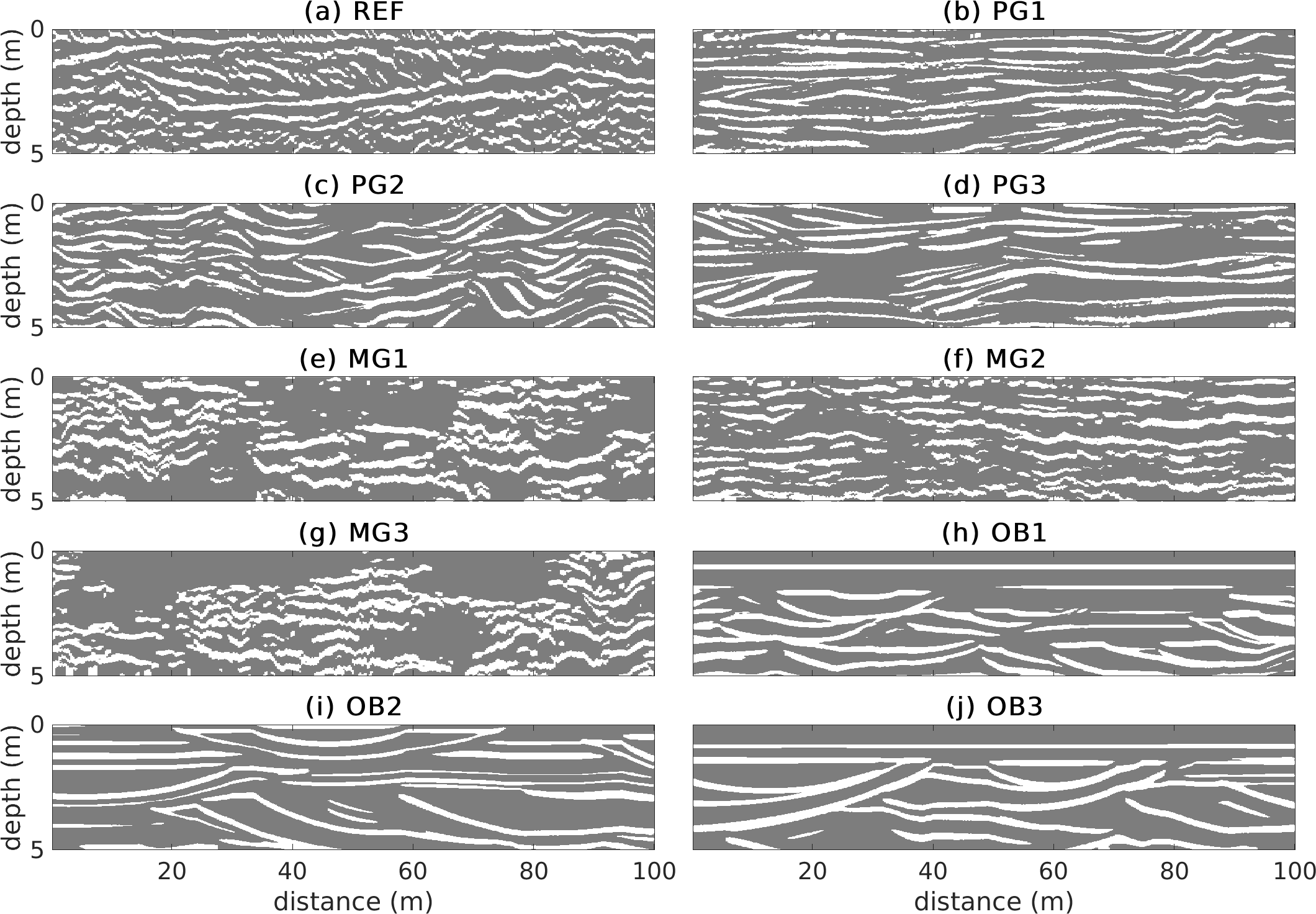}
	\caption{Images obtained after thresholding the example sections represented in Figure \ref{GPfig2}. (a) The binary geophysical image obtained from field data; (b), (c) \& (d) binary geophysical images obtained from pseudo-genetic porosity model realizations for parameter sets PG1, PG2 \& PG3, respectively; (e), (f) \& (g) binary geophysical images simulated from truncated multi-Gaussian porosity model realizations for parameter sets MG1, MG2 \& MG3, respectively; (h), (i) \& (j) binary geophysical images simulated from object-based porosity model realizations for parameter sets OB1, OB2 \& OB3, respectively.}
	\label{GPfig3}
\end{figure}

In order to estimate the distances of each scenario realization to the reference GPR sections REF01, GPR reflection sections are simulated from the corresponding 2D porosity sections. 
%Note that these simulations ignore any 3D effects or small-scale heterogeneity that is not represented in the considered porosity sections. 
The processing steps are: 
\begin{enumerate}
	\item Realization of a facies/porosity model according to a geological conceptual model (scenario) as described in Section \ref{subsecGeologicalPrior}.
    \item Porosity fields are converted into electrical property fields and velocity fields using the model by \cite{pride1994}. The petrophysical parameters (cementation index $m$, and dielectric constant of solid grains $\kappa_s$) are calibrated, such that the mean velocity of the corresponding porosity field is the same as the one used for the field data migration ($\overline{vel}=100 \: $m/\textmu s).
    \item Construction of a perfectly migrated GPR section \citep[following the method developed by][]{irving2010} by convolution of the propagated wavelet with a Primary Reflectivity Section. The propagated wavelet is estimated from field data processing step 5 \citep[according to the method by][]{schmelzbach2015}. % see Figure \ref{GPfig3}
    The Primary Reflectivity Section is derived from the previously obtained velocity model. A simple Gaussian horizontal filter is applied on the convolution result, to account for the Fresnel zone and whose width is determined by the dominant signal wavelength.
    \item To mimic the effect of a constant velocity migration, the GPR reflection section generated with the actual velocities predicted from a porosity model is converted in the time domain before being back transformed into the depth domain using the same mean velocity as the one used in the migration of the field data ($\overline{vel}=100 \: $m/\textmu s), and finally re-interpolated over a regular grid on the vertical axis.
	\item 1D vertical Gaussian filter to slightly smooth the propagated wavelet with the same parameter as the one applied in the processing of the field data.
	\item Automatic gain control to balance signal amplitudes with the same parameters as the one applied in the processing of the field data.
\end{enumerate}
The resulting synthetic GPR sections (Figures \ref{GPfig2}b-j) are thresholded into binary images in the same way as the field data. The binary images resulting from the porosity images in Figures \ref{GPfig1}b-j are given in Figures \ref{GPfig3}b-j.

% Detailed step to estimate the wavelet:
% *** Class GPR ***
%  survey date =  2012-10-02 
%  Reflection, 100MHz,Window length=134.4ns, dz=0.8ns
%  1055 traces,263.5m long
%  > PROCESSING
%    1. coord<-
%    2. crs<-
%    3. dcshift:1-50+mean
%    4. time0<-
%    5. timeCorOffset
%    6. dewow:type=MAD+w=30
%    7. fFilter:f=5,10,170,250+type=bandpass+plotSpec=NULL
%    8. gain:type=power+alpha=1+tcst=10+t0=0+te=200
%    9. gain:type=exp+alpha=0.015+t0=20+te=140
%    10. dewow:type=Gaussian+w=10
%    11. deconv:method=mixed-phase+W=38,125+wtr=5+nf=35+mu=1e-04+shft=1
%  ****************

\section{Results}
\label{secResults}
For each of the three types of geological conceptual models and each of the three corresponding parameter sets (i.e., the nine considered scenarios), we generated 20 porosity realizations. This means, that a total of 180 binary images were available for comparison with the binary reference section REF01 (Figure \ref{GPfig3}a). Wavelet-based, multiple-point histogram, and connectivity distance measures were computed between all possible pairs of field and synthetic binary images as follows.
The wavelet-based distance uses $B=50$ bins and $M=2$ (multi-grid) levels. The MPH-based distance relies on a $5 \times 5$ pixels search-window, $M=3$ multi-grid-levels and on the $O=30$ most frequent patterns. The connectivity-based distance is defined for $A=2$ directions (section length axis $x$ or section depth axis $z$); the investigated distances are limited to half the model dimensions, depending on the axis, and the number of lags is set to $L=25$. 
For each distance type, the distance values are normalized by their maximum. \newline

\begin{figure}[ht!]
	\includegraphics[width=1\textwidth]{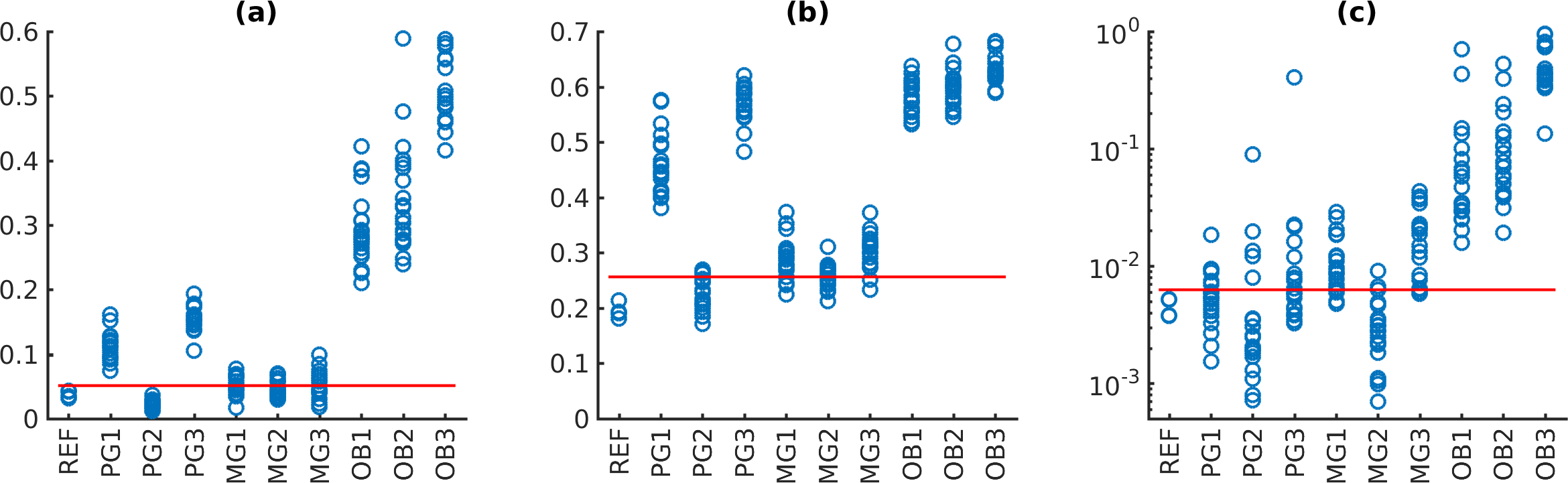}
	\caption{Distance to Tagliamento reference section REF01; plots grouped by scenario for (a) wavelet-based distance, (b) MPH-based distance and (c) connectivity-based distance. REF denotes distances for other binary reference sections (REF02-REF05) with respect to REF01 and the red line corresponds to the acceptance threshold.}
	\label{GPfig4}
\end{figure}

The distances obtained between all binary images and the Tagliamento reference section REF01 are displayed and grouped for each distance type by geological scenario (Figure \ref{GPfig4}). To indicate the internal variability of the distances between the actual field data, the distances between binary reference section REF01 and other binary reference sections (REF02 to REF05) are gathered in a group denoted ``REF''. An acceptance threshold is defined by multiplying by 1.2 the maximum REF distance value. This subjectively-chosen acceptance threshold is used to select realizations whose distances to REF01 is similar to those of the reference sections. \newline

The distances between the primary reference and the scenarios PG2, MG1, MG2, and MG3 are the smallest for the wavelet-based and MPH-based distances. For PG2, the values are close to those of the REF distances, while the MG1, MG2, and MG3 ensembles have mean values that are lower (MG2) or slightly higher (MG1 and MG3) than the acceptance threshold. The connectivity-based distance values are more scattered within each scenario, but most of the PG1, PG2, PG3, and all but one of the MG2 realizations are below the acceptance threshold. The OB1, OB2, and OB3 scenarios are the furthest from the acceptance thresholds for all distance measures considered. \newline

\begin{figure}[ht!]
	\centering
	\includegraphics[width=1\textwidth]{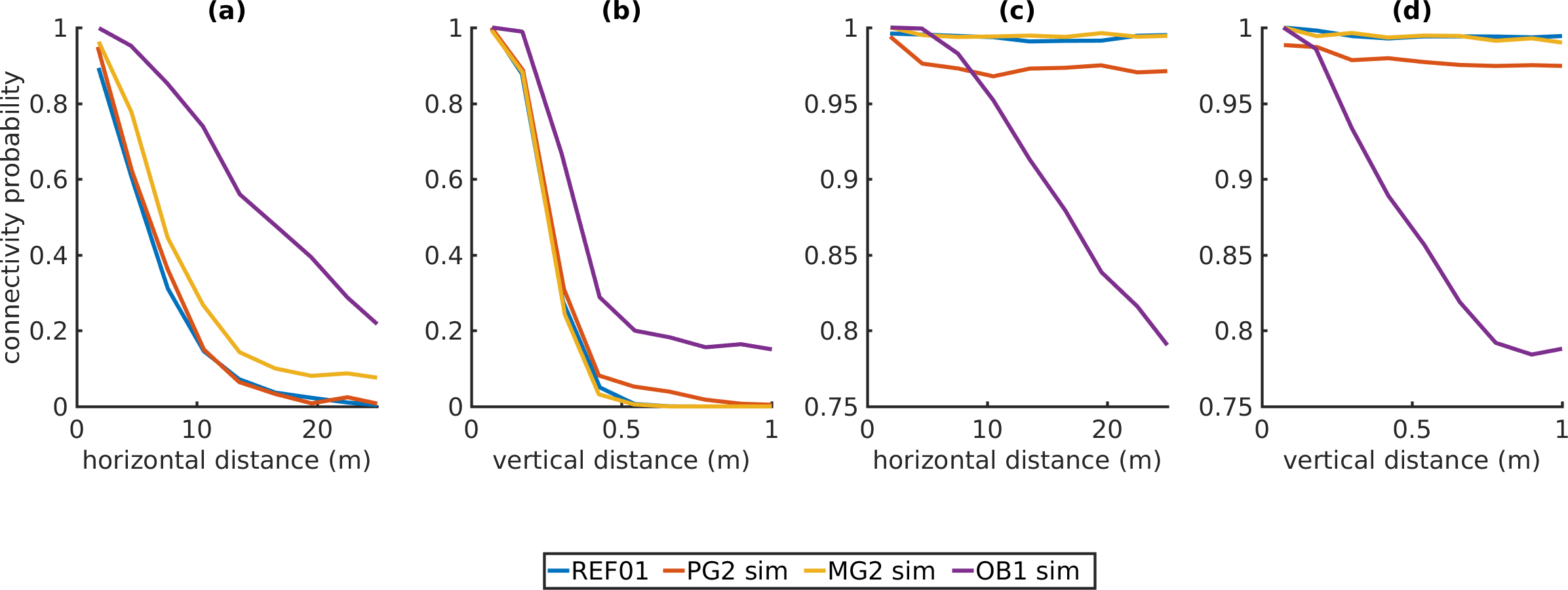}
	\caption{Example of connectivity functions for a selection of binary geophysical images (REF01, PG2 sim, MG2 sim \& OB1 sim from Figure \ref{GPfig3}); (a) horizontal connectivity functions for Class 1 (white) components; (b) horizontal connectivity functions for Class 1 (white) components; (c) horizontal connectivity functions for Class 2 (gray) components; (d) horizontal connectivity functions for Class 2 (gray) components.}
	\label{GPfig4b}
\end{figure}

To better understand the generally-better performance of the PG-family as judged by the connectivity-based distance, we present connectivity functions in Figure \ref{GPfig4b} for some of the sections displayed in Figure \ref{GPfig3}. For the Class 1 components, the horizontal connectivity function (Figure \ref{GPfig4b}a) is best reproduced by PG2, while the connectivity is overestimated for MG2 (by $\approx0.08$) and severely overestimated for OB1 (by 0.1 to 0.5). The vertical connectivity function (Fig. \ref{GPfig4b}b) is best reproduced by MG2, while it is slightly too high for PG2 (at most $\approx0.05$ between 0.4 m and 0.9 m) and far too high for OB1 (by 0.1 to 0.2). For the horizontal and vertical connectivity functions of the Class 2 components (Figure \ref{GPfig4b}c-d), MG2 is found to reproduce them the best, while the connectivity is slightly lower for PG2 (by $\approx -0.02$) and much too small for the OB1 scenario (up to -0.2). \newline

\begin{figure}[ht!]
	\includegraphics[width=1\textwidth]{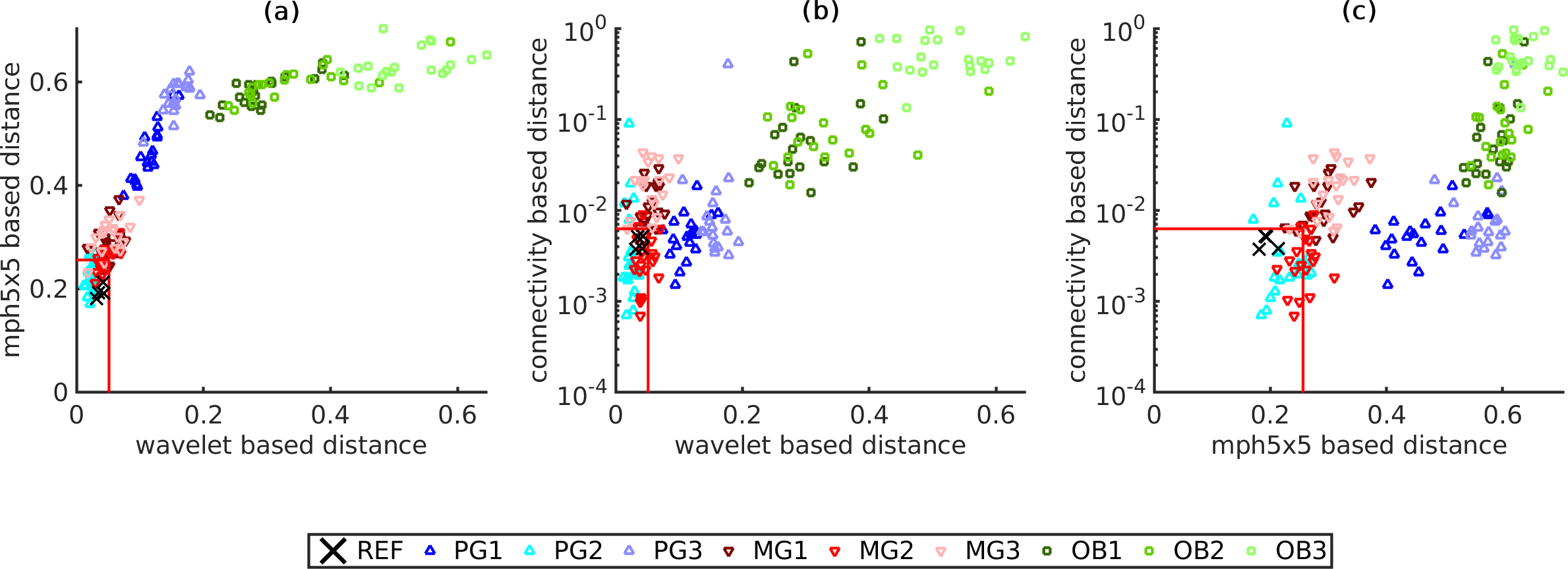}
	\caption{Distance to Tagliamento reference section REF01 visualized as scatter plots grouped by scenario: (a) MPH-based distance as a function of wavelet-based distance; (b) connectivity-based distance as a function of wavelet-based distance; (c) connectivity-based distance as a function of MPH-based distance. REF denotes other binary images processed from additional GPR profiles (REF02-REF05) acquired at the study site and the red line corresponds to the acceptance threshold.}
	\label{GPfig7}
\end{figure}

To highlight the relationships between the distance types, we display three scatter plots (Figure \ref{GPfig7}). A piecewise linear correlation between wavelet-based and multiple-point histogram distances is clearly visible in Figure \ref{GPfig7}a, in which a first segment corresponds to the PG and MG scenarios and a second to the OB scenarios. It also shows the ability of wavelet-based and multiple-point histogram distances to distinguish between the different conceptual models and some scenarios that cluster in different groups. A log-linear relationship with a low correlation between the connectivity- and the wavelet-based distances is visible in Figure \ref{GPfig7}b. A piecewise and scattered log-linear relationship between the connectivity- and the MPH-based distances is visible in Figure \ref{GPfig7}c, in which the first segment  corresponds to the PG and MG scenarios and a second to the OB scenarios.   \newline

For each distance measure considered, the distances for all pairs of images are used to estimate the density of each scenario in the low dimensional space obtained by MDS. To estimate the updated probability of each scenario (Table \ref{GPtab4}), we limit the number of dimensions used such that 95\% of the information is recovered. To achieve this, the two first MDS dimensions are sufficient for the wavelet-based distance, 14 are necessary for the multiple-point-based distance, and three are enough for the connectivity-based distance. For each distance, the estimated probability for a given scenario is proportional to the density of the cloud composed by the scenario realizations at the location of the reference section REF01 in the MDS space.
It informs about the probability that a realization from a scenario is closer to the reference section REF01 relative to the considered scenarios. 
Considering the wavelet-based distance, with an estimated probability of 85.9\%, PG2 is the most probable scenario and MG1 is the second most likely one (14.1\%). For the multiple-point histogram distance, PG2 is by far the most probable scenario (99.9\%). For the case of the connectivity-based distance, MG2 is judged more likely (47.9\%) than PG2 (33.6\%) followed by PG1 (12.7 \%), because it has fewer high and also the smallest distance value.
If we average the probabilities over the types of distances considered, the scenarios that produce realizations that are the closest to the Tagliamento reference section REF01 is PG2, followed by MG2. %\newline

\begin{table}[ht!]
	%\scriptsize
	\small
    \centering
	\begin{tabular}{l c c c c c c c c c}
		\hline
		& \multicolumn{9}{c}{Scenarios}\\
		Distance Based on & PG1 & PG2 & PG3 & MG1 & MG2 & MG3 & OB1 & OB2 & OB3\\
		\hline
		Wavelet Decomposition & - & 85.9 & - & 14.1 & - & - & - & - & -\\
		Multiple-Point Histogram & - & 99.9 & - & - & - & - & 0.1 & - & -\\
		Connectivity Function & 12.7 & 33.6 & 5.5 & - & 47.9 & - & 0.3 & - & -\\
		\hline
		Average & 4.2 & 73.1 & 1.8 & 4.7 & 16.0 & - & 0.1 & - & -\\
		\hline
	\end{tabular}
	\normalsize
	\caption{Estimated scenario probabilities (\%) computed for each type of distance by adaptive kernel smoothing on MDS representations of the simulated and reference sections; values smaller than 0.1\% are not displayed; for each type of distance (row) the probabilities sums to 100\%.}
	\label{GPtab4}
\end{table}

\section{Discussion}
\label{secDiscussion}

\subsection{Geological scenario falsification at the Tagliamento study site}
\label{subsecDiscussionGeol}
By using three different distance metrics quantifying the agreement between field and simulated GPR sections, we reduce geological conceptual model uncertainty at the Tagliamento site. The direct analysis of the distances (Figures \ref{GPfig4} and \ref{GPfig7}) and the estimated probabilities for each type of distance (Table \ref{GPtab4}) led to similar conclusions. For the nine scenarios considered, two are judged significantly more suitable than the others: the PG2 scenario is the most suitable (its realizations are the closest to the Tagliamento reference section REF01), followed by the MG2 scenario. For both the wavelet-based and multiple-point histogram distances, PG2 is the most probable scenario. In the case of the connectivity-based distance, MG2 is judged the most probable scenario, followed by PG2. \newline 

To understand these rankings, let us consider the binary reference section (Figure \ref{GPfig3}a). It reveals that: \crm{1}) Class 1 components (main reflectors) have very small, small, intermediate and long length scales; \crm{2}) Class 1 components are sub-horizontal, and smaller components might present a stronger dip; \crm{3}) the interface between Class 1 and Class 2 components is irregular; \crm{4}) Class 2 components form a connected matrix. For the PG2 scenario realizations (Figure \ref{GPfig3}c), characteristic (\crm{1}), (\crm{2}) \& (\crm{4}) are present, but the interfaces are smooth. For MG2 scenario realizations (Figure \ref{GPfig3}f), characteristic (\crm{1}), (\crm{3}) \& (\crm{4}) are present, but the Class 1 components are too horizontal. The fact that scenarios PG2 and MG2 realizations fulfill three of these four visual criteria might explain the acceptable distance of their realizations to the Tagliamento reference section REF01. For the OB3 scenario realizations, none of the four criteria is fulfilled, which results in high values for all types of distance measures.
From these results, it seems that the representation of cross-stratified deposits, interface roughness, and partially disconnected interfaces are important to reproduce reflection GPR sections at the Tagliamento site. \newline 

None of the proposed OB scenarios match the Tagliamento reference section REF01. We see two main possible explanations: 1) the geometrical parameters of this conceptual model were not well chosen, that is, the size of the scours and the layer thickness might be too large, the density of scours too small, the inner structure of the scours (i.e. inside the truncated semi-ellipsoids) have too thick deposits, when compared to the PG scenarios; or 2) this conceptual model is inherently unsuitable for this site (e.g., interfaces at porosity changes are too clean, without any contour irregularities or apparent roughness when compared to MG scenarios).
This discussion also highlights that identifying the main characteristics present in the reference images and analyzing their absence or presence in images derived from various scenarios may help to propose new conceptual models or scenarios. This suggests a possible iterative process in which initial results are used to guide improvements in the conceptual models considered. \newline

\subsection{Comparison of distance measures}
\label{subsecDiscussionMeas}
We now interpret our results to identify which distance-types are the most suitable. We observe a piecewise linear relationship between the wavelet-based distance and MPH-based distance (Figure \ref{GPfig7}a).
Since there is less overlap between scenarios along the wavelet-based distance axis (Figure \ref{GPfig7}a-b), we conclude that it is more suitable than the MPH-distance to rank geological conceptual models and, to a lesser extent, their most appropriate parameters.
However, the MPH-based distance is also able to classify models according to their geological conceptual model and scenarios (Figure \ref{GPfig7}a and c), but it performs less well than the wavelet-based distance to distinguish scenario PG3 from OB scenarios. This distance appears to better account for local structures (similar patterns between PG and OB) while the wavelet-based distance better accounts for global structures (different shapes: truncated ellipsoids versus the structures of PG models). Indeed, MPS algorithms often have difficulties in reproducing large scale connectivity even when using multi-grid levels \citep{strebelle2002,mariethoz2010,rongier2013}.\newline

The connectivity-based distances differ the most from the other distances and they display a weak log-linear relationship with the wavelet-based distances. They are effective in rejecting the MG1, MG3 and all OB scenarios. The connectivity-based distance clearly separate the OB models from the other model classes (as shown by Figures \ref{GPfig4}c and \ref{GPfig7}b-d) as the reflectors (Class 1) in the OB models are much too connected in length. A corollary of this is that the background Class 2 is less connected (see Figure \ref{GPfig4b}). \newline

Overall, the results suggest that the wavelet-based distance provides the best ability for scenario differentiation. The connectivity-based distance is also interesting because it adds information and helps refining the scenario selection. Moreover, the connectivity-based distance is particularly interesting if the final application includes transport simulations, whose outcome strongly depends on property connectivity.
We also would like to point to previous work \citep{pirot2014b}, which showed that the MPH-based distance is more sensitive to the sign of property contrasts while wavelet-based distance is more sensitive to the magnitude of property contrasts. 
Other fit-for-purpose distances could be considered and global integrative distances, i.e. that combine multiple global distance types could also be useful. \newline  

\subsection{Influence of ranking method and parameter choices}
\label{subsecDiscussionRank}
\begin{figure}[ht!]
	\includegraphics[width=1\textwidth]{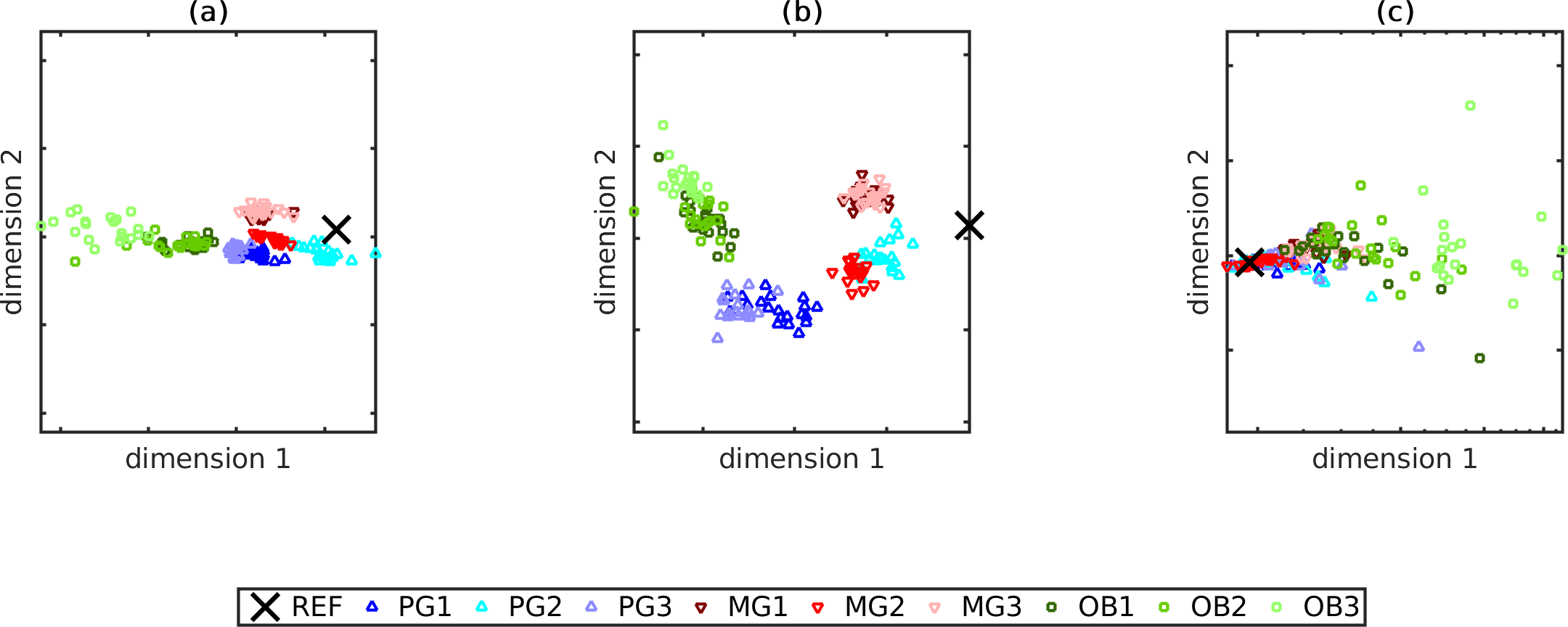}
	\caption{Mapping of the simulated and reference sections in the two first dimensions of the MDS space; (a) for the wavelet-based distance; (b) for the multiple-point based distance; (c) for the connectivity-based distance.}
	\label{GPfig8}
\end{figure}

We have seen that scenario falsification can be performed either by direct analysis of the distances or by estimation of updated probabilities per scenario using MDS followed by adaptive kernel smoothing. On the one hand, direct analysis of distances requires several reference images to define a reasonable acceptance threshold. On the other hand, the estimations of updated probabilities per scenario necessitate the computation of distances for all pairs of images within the ensemble composed of reference and simulated images. Since this cost increases as the square of the number of images, this can become computationally very demanding. Furthermore, rankings and falsifications based directly on distances of scenario probability estimations are relative to the ensemble of considered scenarios. In addition, small distances do not imply that the scenario sections are ``surrounding'' or ``containing'' the reference section in a space mapping the sections (see Figure \ref{GPfig8}). 

Each type of distance requires specific parameter choices. Wavelet-based distances are parameterized by the type of wavelet used (Haar in our case), by the number of decomposition levels (two here) and by the number of bins (50 here). We tested (not shown) the sensitivity to different wavelets (e.g., Daubechies, Coiflets, Symlets, Mexican Hat) and obtained similar results. 
MPH-based distances are parameterized by the pattern size and geometry ($5\times5$ pixels window), the number of multigrid levels (three) and the number of most frequent patterns (30). A number of three \citep{zhang2006,straubhaar2011,straubhaar2013} or four  \citep{strebelle2002,dellArciprete2012} multigrid levels is commonly chosen to generate realizations with tree- or list-based MPS algorithms to capture patterns at multiple scales. The pattern geometry is a basic square which does not favor any anisotropy. The pattern size is kept relatively small to ensure the possibility to encounter similar patterns between images. A smaller pattern size ($3\times3$ pixels window) was tested, but led to similar results. The number of most-frequent patterns is limited to 30 to avoid the comparison of single occurrences that are present only in one of the images. Increasing the number of most-frequent patterns would increase unnecessarily all distances. Decreasing the number of most-frequent patterns would reduce the distances between images. Connectivity-based distances are parameterized by investigated directions and lag width, similarly to the computations of semi-variograms. Here we did not vary these parameters, because the connectivity functions (Figure \ref{GPfig4b}) appears to be well defined. \newline

\subsection{Perspectives}
In the presented case-study, we threshold the reflection GPR sections as part of the data processing (to focus on the main aspects of the reflectors) and limit our comparison to binary geophysical images. One could also apply the proposed methodology to continuous images. It would then be straightforward to compute a distance based on wavelet decomposition. However, multiple-point histograms and connectivity functions as defined in Section \ref{secDistances} are applicable to discrete domain images only. One solution is to threshold the continuous images, as we did here, in a reasonable number of classes, to retrieve and compare the most important features from the images. Of course, this implies some qualitative assessment of which features are the most important ones, depending on the target of modeling. Another possibility is to adapt the definition of the multiple-point histogram and of the connectivity functions, such that they can be applied to continuous images. For instance, we could rely on the definition of distances between continuous patterns \citep{mariethoz2010} and on the identification of pattern clusters to build a multiple-point histogram between continuous images; the pattern clusters could be referred to as the histogram bins, and a pattern could be assigned to the closest bin/cluster; it would though depend on the number of clusters and how they are identified. Regarding the connectivity-based distance, the simplest option would be to define the connectivity as a function of the threshold \citep{meerschman2013,renard2013}, as the probability that two pixels are both above or both below a threshold. \newline

While migrated GPR sections obtained from field data are somehow affected by 3D geological heterogeneities, the simulation of GPR reflection sections is performed from 2D porosity sections and does not account for 3D effects. The binary thresholding is a way to focus on the reflections of interest and to reduce the impact caused by the inherent limitations of the forward modeling, such as considering 3D effects negligible, grid resolution, different coupling effect at the surface, non-horizontal antennae at all times due to small changes in topography, approximations of the propagated wave, estimation of the attenuation with depth, etc. A consequence is that we loose some information about porosity contrasts. Here, it allows to simulate GPR reflection sections very efficiently, and thus to perform conceptual model uncertainty reduction. A way to account for 3D effects would be to perform full-waveform GPR modeling over 3D porosity models. It would tremendously increase the computational requirements, and consequently would make conceptual model selection and falsification very costly. However, characterizing the effects of such model simplifications could improve (quantitatively) our understanding of GPR modeling errors and allows us to mitigate potential bias effects.

\section{Conclusions}
\label{secConclusion}

We have demonstrated how global distances (defined from wavelet decomposition, multiple-point histograms and connectivity analysis) between geophysical images allowed us to falsify seven out of nine considered geological scenarios at the Tagliamento site. By considering GPR sections from the Tagliamento aquifer, we find that cross-stratified deposits and irregular property interfaces are important features to reproduce. An underlying assumption of this work is that the results obtained by model comparison with geophysical data are informative for subsurface flow and transport. This assertion should be tested by tracer tests, that are, up to date, not available at the Tagliamento site. We have found that scenario falsification can be performed either by direct analysis of the distances or by estimation of updated probabilities. Direct analysis is faster, more intuitive and rely on the definition of a subjective acceptance threshold that is informed by the magnitude of distances computed between several reference sections. Computation of scenario probabilities using MDS to map geophysical images as points in a lower dimensional space, followed by adaptive kernel smoothing to estimate scenario probabilities, is more advanced and requires more computing resources.
The use of distance comparisons in geophysics also serves to select new parameter sets or to propose new geological conceptual models, in order to further close the gap between simulated sections obtained from an initial set of scenarios and the reference sections. This approach can be used for any type of geophysical images, as long as the geophysical modeling and processing step can be simulated in an effective and trustworthy way. The most convenient distance of those considered is the wavelet-based distance, which is the fastest to compute and it offers the best clustering of scenarios. The connectivity-based distance add further independent information and should be considered if structure connectivity is expected to have an impact on the prediction variables of interest.
This work proposes a way forward to use uninterpreted GPR data, in contrast to hand-drawn geological deposit interpretation, for quantitative subsurface characterization. 

\section*{Acknowledgments}
The authors thank two anonymous reviewers for their constructive comments, and C\'{e}line Scheidt for sharing her MATLAB code to compute updated probabilities. The other MATLAB codes and the synthetic data used to test the proposed method are available upon request at \href{mailto:guillaume.pirot@unil.ch}{guillaume.pirot@unil.ch} (note that multiple-point histograms computations in that code require the \emph{Impala} software, whose academic license is available upon request at \href{mailto:philippe.renard@unine.ch}{philippe.renard@unine.ch}). The software package to simulate the object-based model is freely available at \href{https://github.com/emanuelhuber/CBRDM}{github.com/emanuelhuber/CBRDM}. The GPR data are available upon request at \href{mailto:emanuel.huber@alumni.ethz.ch}{emanuel.huber@alumni.ethz.ch}. 

%\bibliography{../../../articles/citations/referencesDB}
\bibliography{referencesDB}

% \appendix
% \section{Detailed plots of the connectivity functions}

% \begin{landscape}
% % \vspace*{-1.0cm}
% % \hspace*{-4.0cm}
% \label{appendixConnectivityDetails}
% \begin{figure}[ht!]
% 	\includegraphics[width=24cm]{draft_fig5}
% %     \includegraphics[width=1\textwidth]{draft_fig5}
% 	\caption{Connectivity functions for class 1 of binary sections; first three columns: horizontal connectivity; last three columns: vertical connectivity; first row for PG models; second row for MG models; third row for OB models.}
% 	\label{GPfig5}
% \end{figure}
% \begin{figure}[ht!]
% 	\includegraphics[width=24cm]{draft_fig6}
% % 	\includegraphics[width=1\textwidth]{draft_fig6}
% 	\caption{Connectivity functions for class 2 of binary sections; first three columns: horizontal connectivity; last three columns: vertical connectivity; first row for PG models; second row for MG models; third row for OB models.}
% 	\label{GPfig6}
% \end{figure}
% \end{landscape}
% \clearpage

\end{document}